\documentclass[pra,showpacs,amsmath,amssymb, twocolumn, 10pt]{revtex4}

\usepackage{amsthm}
\usepackage{dcolumn}
\usepackage{bm}
\usepackage{graphicx}

\begin{document}

\newtheorem{corollary}{Corollary}
\newtheorem{definition}{Definition}
\newtheorem{example}{Example}
\newtheorem{lemma}{Lemma}
\newtheorem{proposition}{Proposition}
\newtheorem{theorem}{Theorem}
\newtheorem{fact}{Fact}
\newtheorem{property}{Property}
\newcommand{\bra}[1]{\langle #1|}
\newcommand{\ket}[1]{|#1\rangle}
\newcommand{\braket}[3]{\langle #1|#2|#3\rangle}
\newcommand{\ip}[2]{\langle #1|#2\rangle}
\newcommand{\op}[2]{|#1\rangle \langle #2|}

\newcommand{\tr}{{\rm tr}}
\newcommand {\E } {{\mathcal{E}}}
\newcommand {\F } {{\mathcal{F}}}
\newcommand {\diag } {{\rm diag}}

\title{Exact Quantum Search by Parallel Unitary Discrimination Schemes}

\author{Xiaodi Wu$^{1,2}$}
\email{wuxd04@mails.tsinghua.edu.cn}

\author{Runyao Duan$^{1}$}
\email{dry@tsinghua.edu.cn}

\affiliation{$^1$State Key Laboratory of Intelligent Technology and
Systems, Tsinghua National Laboratory for Information Science and
Technology, Department of Computer Science and Technology, Tsinghua
University, Beijing 100084, China
\\
$^2$Department of Physics, Tsinghua University, Beijing 100084,
China}

\begin{abstract}
We study the unsorted database search problem with items $N$ from
the viewpoint of unitary discrimination. Instead of considering the
famous $O(\sqrt{N})$ Grover's the bounded-error algorithm for the
original problem, we seek for the results about the exact
algorithms, i.e. the ones succeed with certainty. Under the standard
oracle model $\sum_j (-1)^{\delta_{\tau j}}|j\rangle\langle j|$, we
demonstrate a tight lower bound $\frac{2}{3}N+o(N)$ of the number of
queries for any parallel scheme with unentangled input states. With
the assistance of entanglement, we obtain a general lower bound
$\frac{1}{2}(N-\sqrt{N})$. We provide concrete examples to
illustrate our results. In particular, we show that the case of
$N=6$ can be solved exactly with only two queries by using a
bipartite entangled input state. Our results indicate that in the
standard oracle model the complexity of exact quantum search with
one unique solution can be strictly less than that of the
calculation of OR function.
\end{abstract}

\pacs{03.67.-a,03.67.Lx,03.65.Ud}

\maketitle

\section{Introduction}~\label{sec:intro}
Quantum computing is more powerful than classical computing due to
many peculiar features of quantum mechanics such as superposition
and entanglement. Although it is still unclear whether quantum
computer can efficiently solve NP-complete problem, there do exist
some problems for which quantum algorithms outperform any known
classical algorithms. Outstanding instances include the Shor's
algorithm for factoring large integers \cite{SHO94}, and the
Grover's algorithm for searching a specific element in an unsorted
database \cite{GRO97}.

\emph{The Unsorted Database Search Problem} can be formulated as
follows. Suppose we have a database whose elements are labeled from
$1$ to $N$, and suppose we have a function $f$: $\{1,\cdots,
N\}\rightarrow\{0,1\}$. Assume there is a unique element $x_0$ in
the database such that $f(x_0)=1$. The goal of the problem is to
figure out $x_0$ with the minimum number of calculations of $f$.

Usually, we treat the function $f$ as a black-box or an oracle. We
use a query to the oracle to get the value $f(x_i)$ when the input
to the oracle is $x_i$. In classical computing, the minimum number
of queries to the oracle is used to measure the complexity of the
original problem. In quantum computing, we have the counterpart of
the classical oracle, namely the quantum oracle. A standard quantum
oracle $O_f$ for a boolean function $f$ on $\{1,\cdots, N\}$ is
defined as follows:
\begin{equation}\label{oracle-f}
O_f\ket{x}\ket{y}=\ket{x}\ket{y\oplus f(x)},
\end{equation}
where $\{\ket{x}:1\leq x\leq N\}$ is an orthonormal basis for the
principal quantum system of interest, and $\{\ket{y}:y=0,1\}$ is an
orthonormal basis for the auxiliary qubit which is used to store the
result of query. If we prepare the auxiliary system in state
$\ket{-}=\frac{1}{\sqrt{2}}(\ket{0}-\ket{1})$, the action of $O_f$
on the principal system can be simplified to the following form:
\begin{equation}\label{oracle-simplified}
O_{f}=\sum_{j=1}^N (-1)^{f(j)}\op{j}{j}.
\end{equation}

\begin{figure}[ht]
  \centering
  \includegraphics[scale=0.65]{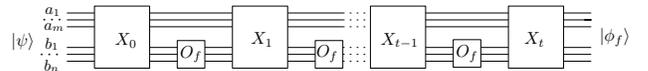}
  \caption{Illustration of Quantum Network with $t$ queries.}
  \label{QC}
\end{figure}

A \emph{general quantum network} with $t$ queries can be visualized
in Fig. \ref{QC}, where $O_f$ stands for the quantum oracle, and
each $X_i$ is a known unitary operation inserted between two
successive queries of the oracle. The input to the network is
$\ket{\psi}$ with $m+n$ qubits where the first $m$ qubits represent
the auxiliary qubits, and the last $n$ qubits are the qubits
relative to the oracle. $X_i$ will affect on $m+n$ qubits while
$O_f$ will only affect the last $n$ qubits. The computation is
completed with a measurement on the final output state $\ket{\phi}$.

The computation with oracles has the fixed $\{X_0,\dots,X_t\}$ and
the same input state $\ket{\psi}$. Thus, the output state of the
computation $\ket{\phi}$ relies only on the quantum oracle $O_f$,
namely the function $f$. As introduced in Eq. (\ref{oracle-f}), the
function $f$ will determine $O_f$ and thus the output state
$\ket{\phi}$, which is named $\ket{\phi_f}$. Different function
$f_i$ will result in different output state $ \ket{\phi_i}$. For an
exact algorithm, we need to distinguish the set of output states $\{
\ket{\phi_i} \}$ with \emph{certainty}. In other words, any two
states $\ket{\phi_i}$ and $\ket{\phi_j}$ such that $i\neq j$ should
be orthogonal. If this orthogonality condition cannot be satisfied,
the algorithm will fail to distinguish all the possible output
states. In such case, $\ip{\phi_j}{\phi_i} \neq 0$ ($i\neq j$), the
algorithm may output $i$ if the actual result is $j$ and vice versa
with some positive probability. If the probability satisfies certain
requirements, we call such algorithm a \emph{bounded-error}
algorithm. It should be noted that the number of queries used in
this network is the measure of the computational complexity.

Grover~\cite{GRO97} invented an efficient algorithm to the unsorted
database search problem using the quantum network above. More
precisely, in his scheme, $m=0$, namely no auxiliary qubits. All
$X_i$ and $O_f$ will affect on the last $n$ qubits. Using the
notations above, we have:
\begin{equation} \label{equ:evolution}
\ket{\phi_f}=X_tO_fX_{t-1}O_f\cdots X_1O_fX_0\ket{\psi},
\end{equation}
Furthermore, $O_f$ in this problem has the form below:
\begin{equation}\label{oracle-simplified}
O_{f}=\sum_{j=1}^N (-1)^{\delta_{x_0j}}\op{j}{j},
\end{equation}
where $\delta_{ab}$ or $\delta_a^b$ equals to $1$ if $a=b$ and
equals to 0 otherwise, and $x_0$ is the unique $x$ such that
$f(x)=1$. With a careful choice of $\{X_i\}$ and input state
$\ket{\psi}$, Grover obtained a \emph{bounded-error} algorithm using
only $O(\sqrt{N})$ queries. Pioneering work in Ref. \cite{BBBV97}
presented a lower bound of $\Omega(\sqrt{N})$
\footnote{$\Omega(f(n))$ means $\geq cf(n)$ where $c$ is a positive
constant.}. Combined with Zalka's work~\cite{Zalka99}, it
immediately implies the Grover's algorithm is optimal for
bounded-error setting. However, the original Grover's algorithm
succeeds with certainty only when $N=4$.

For exact algorithm, under the oracle model $\sum_j
(-1)^{\delta_{\tau j}}|j\rangle\langle j|$, people obtained the
complexity of the decision version of quantum search problem with
multi-solution, where they treated the decision version problem as
the calculation of function OR \cite{BBCM+98}. In such situation,
$\Omega(N)$ queries is required to get the answer with certainty.
However, in this paper, we care about the quantum search problem
with unique-solution. Thus, the complexity is no more than
$\Omega(N)$. In classical computing, $N-1$ is necessary for exact
algorithm. As the generalization of Grover's quantum search
algorithm, quantum amplitude amplification was proposed
in\cite{BHMT00}. Later, arbitrary phase concept was introduced
\cite{Hoyer01,longctp99,longpla99} under the modified oracle model
$\sum_j e^{\textbf{i}\theta\delta_{\tau j}}|j\rangle\langle j|$,
where $\textbf{i}=\sqrt{-1}$. H{\o}yer \cite{Hoyer01} and Long
\cite{LON01} further employed such concept and successfully found an
exact algorithm using $O(\sqrt{N})$ queries to solve the quantum
search problem. Also, the computing problem of boolean function OR
is thought to have a close relation with the unsorted database
search problem \cite{WL07}. Quantum lower bounds for such boolean
functions have been thoroughly discussed in~Ref. \cite{BBCM+98}.

\begin{figure}[ht]
  \centering
  \includegraphics[scale=0.65]{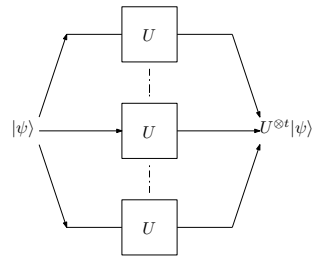}
  \caption{A parallel discrimination scheme $U^{\otimes t}$ with $t$ runs.}
  \label{PARALLEL}
\end{figure}

Another interesting problem which has received considerable
attention is the discrimination of unitary operations. Suppose that
we have an unknown unitary operation $U$ which is secretely chosen
from a set of pre-specified unitary operations, say, $\{U_1,\cdots,
U_N\}$. Our task is to decide the real identity of this unitary,
i.e. the index of $U$. To do this, we employ a similar network as
Fig. \ref{QC} with the change that replacing $O_f$ by $U$. In
particular, we call the network a {\it parallel scheme} if the
network is reduced to the form of $U^{\otimes t}$. See Fig.
\ref{PARALLEL}.

\begin{figure}[ht]
  \centering
  \includegraphics[scale=0.5]{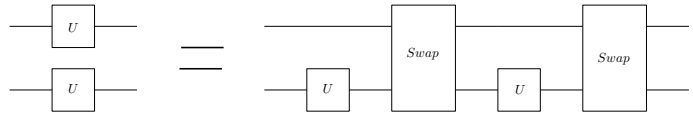}
  \caption{Parallel scheme as a special case of general quantum network.}
  \label{EQUI}
\end{figure}

To verify that a parallel scheme is a special case of quantum
network, one only needs to notice the identity
$$U^{\otimes 2}=(I\otimes U)S(I\otimes U)S^\dagger,$$
where $S$ is the swap operation, i.e.,
$S\ket{\psi}\ket{\phi}=\ket{\phi}\ket{\psi}$ for any $\ket{\psi}$
and $\ket{\phi}$. See Fig. \ref{EQUI} for an intuitive illustration.

When the unknown unitary is $U_k$, we obtain $\ket{\phi_k}$ as the
output of the discrimination network. To perfectly distinguish
between $\{U_k\}$, we need to distinguish among $\{\ket{\phi_k}\}$.
So all these states should be mutually orthogonal. The number of $U$
appearing in this network, or the number of runs of $U$ is the cost
of the network. Due to its special structure, a parallel scheme can
accomplish the discrimination with a single step when a large number
copies of $U$ are available.

Unlike the discrimination of nonorthogonal states, which is
impossible even arbitrarily large but finite number of copies are
available, we can always discriminate any finite set of unitary
operations with certainty using some quantum network. Actually it is
possible to achieve a perfect discrimination between unitary
operations by a parallel scheme \cite{AC01, DLP01} with the
assistance of a multipartite entangled state as input.
Interestingly, it was further proven that the entangled input state
is not necessary by employing a sequential scheme instead of a
parallel one \cite{DFY07}. An analytical expression for the minimal
number of runs needed for a perfect discrimination between two
unitary operations using general quantum network was also obtained
in Ref. \cite{DFY07}. Very recently it was shown that any two
multipartite unitary operations can always be perfectly
distinguished by local operations and classical communication
\cite{DFY07b}.

Since unitary operations are the natural generalization of classical
boolean functions in the quantum computing setting, it is obvious
that the problem of distinguishing unitary operations is actually
the quantum counterpart of oracle identification problem in
classical computing. Many works have been done in order to sharpen
our understanding of both classical and quantum oracle
identification problem \cite{AIKM+04,AIKR+06}.

The purpose of the paper is to study the unsorted database search
problem from the viewpoint of unitary operation discrimination. It
is easy to see the original problem is equivalent to find which
$f_i$ is currently in use, namely an oracle identification problem,
which we name it \emph{Grover's Oracle Identification Problem}. We
want to solve the problem with oracles in
Eq.~(\ref{oracle-simplified}) with \emph{certainty} using a parallel
discrimination scheme. For unsorted database problem, the candidate
set is $\{ O_{i}: 1\leq i \leq N\} $ where $O_{i}$ is in the form of
Eq.~(\ref{oracle-simplified}) and $N$ is the size of the database in
the problem.


The known quantum lower bound for the problem in discussion is
$\Omega(N)$. It is somewhat surprising that we can solve the
Grover's oracle identification problem by a parallel discrimination
scheme without entanglement using at most $\frac{2}{3}N+2$ queries,
which is strikingly different from the classical setting, where
$N-1$ queries are necessary. We further show that such a scheme is
optimal for any parallel scheme without entanglement. We also find
that entanglement may reduce the number of queries and thus improve
the efficiency of discrimination. In particular, a lower bound
$\frac{1}{2}(N-\sqrt{N})$ for the discrimination with entanglement
is obtained. Most interestingly, we show that two queries are
sufficient for a perfect discrimination for $N=5$ without use of
entanglement, and are still sufficient for a perfect discrimination
for $N=6$ if an entangled input state is allowed. It is also worth
noting that in our proofs we have extensively employed the
techniques from the graph theory and combinatorics.  We hope these
proof techniques may be useful for other problems in quantum
computation and quantum information.

\section{Parallel Discrimination Scheme for Exact Quantum Search} ~\label{sec:parascheme}
A parallel discrimination scheme which is visualized in
Fig.~\ref{PARALLEL} uses the network $U^{\otimes t}$ where $U$ is
the unitary operation to identify from a candidate set and $t$ is
the number of the copies, namely the complexity of the scheme. More
precisely, in \emph{Grover's Oracle Identification Problem}, the
candidate set is $\{f_i\}$ where $f_i=\sum_j
(-1)^{\delta_{ij}}|j\rangle\langle j|$, the set of possible output
states of the network is $\{f_i^{\otimes t}|\psi\rangle\}$. Since
the algorithm is exact, these output states should be orthogonal to
each other. That is,

\textbf{Discrimination Condition}: For any $1\leq i<j\leq N$,
\begin{equation}\label{eqn:discondition}
   \langle\psi|{(f_i^{\otimes t})}^{\dagger}{(f_j^{\otimes
t})}|\psi\rangle=0,
\end{equation}
where
\begin{equation}\label{eqn:expandoracle}
  {(f_i^{\otimes t})}^{\dagger}{(f_j^{\otimes
  t})}=\sum_{\vec{a}} (-1)^{\tau(\vec{a})}
  |\vec{a}\rangle\langle \vec{a}|,
\end{equation}
where $\vec{a}=a_1\cdots a_t$ and
\begin{equation}\label{eqn:sgnoracle}
 \tau(\vec{a})=\sum_{k=1}^t (\delta_{a_k}^i+\delta_{a_k}^j)\mod 2.
\end{equation}
In the following, we will first show the case with only one-copy
state. Using one-copy state as a product state block, we obtain a
product state discrimination scheme and prove its optimality.
Finally, we deal with the scheme with entanglement and give
examples.

\subsection{A key lemma}
Suppose $\ket{\psi}$ is an input state in $\mathcal{H}_N$. We say
that $\ket{\psi}$ can discriminate a pair $(i,j)$ if $f_i\ket{\psi}$
and $f_j\ket{\psi}$ are orthogonal. Note that here a pair $(i,j)$ is
just an abbreviation for $\{i,j\}$. Let $S_{\ket{\psi}}$ represent
the pairs that can be discriminated by $\ket{\psi}$, i.e.,
\begin{equation}\label{S-set}
S_{\ket{\psi}}=\{(i,j):\braket{\psi}{f_i^\dagger f_j}{\psi}=0\}.
\end{equation}
With the above notation, we can visualize the discrimination power
of $\ket{\psi}$ by the \emph{discrimination graph} defined as
follows.

\textbf{Discrimination Graph}: an undirected graph
$G_{\ket{\psi}}=(V,E)$ with vertex set $V=\{1,\cdots, N\}$ and edge
set $E=S_{\ket{\psi}}$. We find that such a graph representation may
be helpful in understanding the following arguments.

Assume that $\ket{\psi}$ is of the form $\sum_i p_i|i\rangle$, where
$\sum_i |p_i|^2=1$. By Eq. (\ref{eqn:discondition}), we can easily
see that $\ket{\psi}$ can discriminate a pair $(i,j)$ if and only if
\begin{equation}\label{eqn:t1discondition}
   |p_i|^2+|p_j|^2=\frac{1}{2}.
\end{equation}
Three kinds of interesting states which satisfy the above condition
for certain pairs $(i,j)$ are as follows:
\begin{itemize}
  \item $|\psi\rangle=\frac{1}{2}(|a\rangle+|b\rangle+|c\rangle+|d\rangle)$
  with distinct $\{a,b,c,d\}$ can discriminate any pair in $\{a,b,c,d\}$.
  We denote such states as $K_4\{a,b,c,d\}$ or $K_4$ in short. See
  Fig.\ref{k4}.
\begin{figure}[ht]
  \centering
  \includegraphics[scale=0.6]{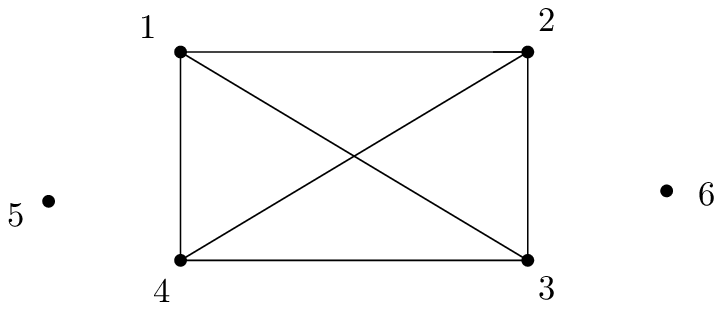}
  \caption{Illustration of $S_{K_4\{1,2,3,4\}}$, $N=6$.}
  \label{k4}
\end{figure}
  \item $|\psi\rangle=\frac{1}{\sqrt{2}}(|i\rangle+|j\rangle)$ with distinct $i,j$
  can discriminate any pair $(i,k)$ or $(j,k)$ where $k\notin\{i,j\}$, however the pair $(i,j)$ cannot be discriminated.
  We denote such states as $<i,j>$. See Fig. \ref{S12}.
\begin{figure}[ht]
  \centering
  \includegraphics[scale=0.6]{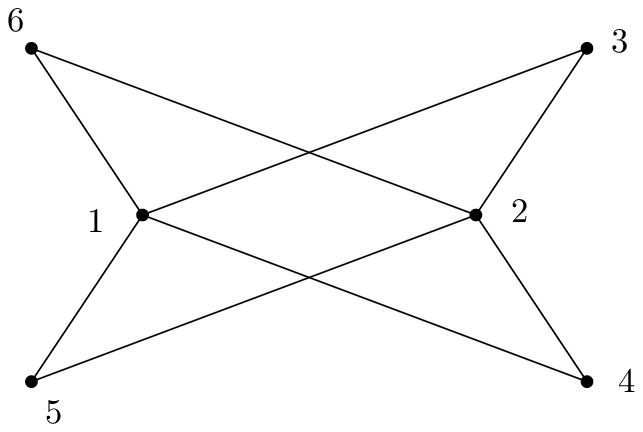}
  \caption{Illustration of $S_{<1,2>}$, $N=6$.}
  \label{S12}
\end{figure}

  \item $|\psi\rangle=a|i\rangle+b\sum_{j\neq i}|j\rangle$, where
  $a=\sqrt{\frac{N-3}{2(N-2)}}$ and $b=\sqrt{\frac{1}{2(N-2)}}$ for $N\geq
  3$. This state discriminates all the pairs $(i,c)$ where $c\neq
  i$. We denote such states as $E(i)$. See Fig. \ref{E(1)}.
  \begin{figure}[ht]
  \centering
  \includegraphics[scale=0.6]{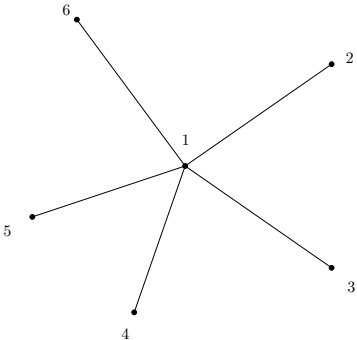}
  \caption{Illustration of $S_{E(1)}$, $N=6$.}
  \label{E(1)}
\end{figure}

\end{itemize}

We would like to choose input state such that it can discriminate as
many as possible pairs. A state $\ket{\psi}$ is said to be trivial
if it cannot discriminate any pair, namely
$S_{\ket{\psi}}=\emptyset$. Trivial states are useless for
discrimination. We only focus on nontrivial states. Surprisingly,
any nontrivial state $\ket{\varphi}$ can be replaced by at least one
of the three states above say $\ket{\psi}$ such that
$S_{\ket{\varphi}}\subseteq S_{\ket{\psi}}$. This is crucial as it
considerably simplifies the original identification problem. We
formulate it as follows.
\begin{lemma}\label{lemma1}\upshape
If any non-trivial one-copy input state $|\psi\rangle=\sum_i
p_i|i\rangle$ is not $K_4$ or $<i,j>$ or $E(i)$, we can always use
one of  $K_4$, $<i,j>$, or $E(i)$ to replace it such that
$S_{\ket{\psi}}\subseteq S_{K_4}$ or $S_{<i,j>}$ or $S_{E(i)}$.
\end{lemma}
\textbf{Proof.} For a non-trivial state $|\psi\rangle=\sum_i
p_i|i\rangle $ which is not $K_4$ or $<i,j>$ or $E(i)$, it must be
able to discriminate at least one pair. There are two cases we need
to consider:

  Case 1. If the pairs discriminated by $\ket{\psi}$ are not mutually
    disjointed, say the pair $(i,j)$ is one of them, then the other
    pairs (if exist) can only be of the form $(i,k)$ or $(j,k')$.
    Because for the pair $(i,k)$ or $(j,k')$ , the other pairs should also not be
    disjointed with it, there are only three cases which satisfy the
    assumption. The first case is that only one pair $(i,j)$ can be discriminated.
    The second case is that there are three pairs $(i,j)$,$(i,k)$ and $(j,k)$ which can be discriminated. In
    both cases, $K_4\{i,j,k,l\}$ can replace $\ket{\psi}$. The
    third case is $\ket{\psi}$ discriminates the pairs in $\{(i,c_k),c\neq i\}$
    or $\{(c,j),c\neq j\}$, then we can use $E(i)$ or $E(j)$
    respectively to replace $\ket{\psi}$.

   Case 2. If there are two disjointed pairs $(a,b),(c,d)$, due to
    Eq.~\ref{eqn:t1discondition}, we have
      $|p_a|^2+|p_b|^2 = \frac{1}{2}$ and $|p_c|^2+|p_d|^2 = \frac{1}{2}$,
    namely $|p_a|^2+|p_b|^2+|p_c|^2+|p_d|^2=1$, which implies other $p_e=0$.
    In order to satisfy the condition above, at least one variable
    in each equation is non-zero. If there are only $2$ (say $p_a,p_c$) or $4$ variables
    are non-zero, then we can use $<a,c>$ or $K_4$ to replace $\ket{\psi}$
    respectively. Otherwise, we have $p_a,p_b,p_c$ non-zero,
    $<a,c>$ is also able to replace $\ket{\psi}$.

In both cases  $\ket{\psi}$ can be replaced successfully. That
completes the proof. \hfill $\blacksquare$

As a direct consequence of Lemma \ref{lemma1}, the power to
discriminate pairs by one-copy state cannot exceed the power of
$K_4$ or $<i,j>$ or $E(i)$. Thus, one-copy state isn't adequate for
discrimination when $N\geq 5$. It is easy to verify that there is no
discrimination scheme for the cases of $N=2, 3$. Only for $N=4$ we
have a discrimination scheme with using one single copy. However, if
we choose another oracle model say the auxiliary qubit
$|y\rangle=|0\rangle$, we can discriminate the case $N=2$ with input
state $|\psi\rangle=\frac{1}{\sqrt{2}}(|1\rangle+|2\rangle)$.
Furthermore, we can see in that oracle model, one-copy state can
only discriminate one pair which is much less powerful than the one
in our approach.

\subsection{Unentangled Discrimination Scheme}
In the following we shall present a scheme of discrimination without
any use of entanglement. In such scheme, the input state of $t$-copy
network must be of the form
\begin{equation}\label{eqn:entanglefreestate}
   |\psi\rangle=|\psi_1\rangle\otimes|\psi_2\rangle\otimes...\otimes|\psi_t\rangle
\end{equation}

Substitute the input state to the condition~\ref{eqn:discondition},
for any pair $(i,j)$,
$\langle\psi_1|f_i^{\dagger}f_j|\psi_1\rangle\langle\psi_2|
f_i^{\dagger}f_j|\psi_2\rangle...\langle\psi_t|f_i^{\dagger}f_j|\psi_t\rangle=0$.
Thus, at least one block
$\langle\psi_n|f_i^{\dagger}f_j|\psi_n\rangle$ must be 0 in order to
satisfy the equation. Namely the pair $(i,j)$ is discriminated by at
least one $|\psi_n\rangle$.

In order to satisfy all the pairs in
condition~\ref{eqn:discondition}, any pair $(i,j)$ must be
discriminated by at least one $|\psi_n\rangle$ in
~\ref{eqn:entanglefreestate}. A \emph{discrimination scheme} is
defined as a scheme to discriminate all the pairs in $\{1,...,N\}$
which are denoted as the set $S_{all}$ or $K_N$. We define the
discrimination scheme with input state in the form of
$|\psi\rangle=|\psi_1\rangle\otimes|\psi_2\rangle\otimes\cdots\otimes|\psi_t\rangle$
as \emph{product discrimination scheme} or \emph{unentangled
discrimination scheme}. Then for such scheme, it is required that
$\cup_{n=1}^t S_{\ket{\psi_n}}=S_{all}$.

It is interesting to find out the minimal $t$ for any unentangled
discrimination scheme. Any block
$\langle\psi_n|f_i^{\dagger}f_j|\psi_n\rangle$ which can not
discriminate any pair namely trivial will not belong to the scheme
due to the minimum copies requirement. We propose a product
discrimination scheme with $\frac{2}{3}N+1$ queries.

\textbf{Discrimination Scheme:} we construct the scheme only with
$<i,j>$ state. We divide all the $N$ elements into groups, where
each group contains 3 elements. Say the group of $\{1,2,3\}$, we use
$<1,2>$ and $<1,3>$ to discriminate all the pairs one of whose
elements is in $\{1,2,3\}$. It is easy to see we can use the same
scheme for every group. If $N$ elements cannot be divided into
groups of 3 elements exactly, there is a incomplete group of 1 or 2
elements say $\{N\}$ or $\{N-1,N\}$. We can use extra $<1,N>$ or
$<1,N-1>$ to discriminate the pairs which have one elements in the
incomplete group.
It is easy to verify that such a scheme is valid and use only at
most $\frac{2}{3}N+1$ copies.

However, it is surprising to see that such a simple scheme reaches
the lower bound of product discrimination scheme. First, we will
show it is sufficient to consider only 3 types states to simply our
proof. A replacement of a block $\ket{\varphi}$ by $\ket{\psi}$ in
\emph{unentangled discrimination scheme } is \emph{valid} if
$S_{\ket{\varphi}}\subseteq S_{\ket{\psi}}$. It is easy to see that
such a replacement won't diminish $\cup_{n=1}^{t} S_{\ket{\psi_n}}$
which guarantees the scheme after the replacement is also a  valid
discrimination scheme if it is before.

\begin{lemma}\label{lemma2}\upshape
For any \emph{unentangled discrimination scheme}, if any block
$|\psi_n\rangle\ $ is not $K_4$ or $<i,j>$ or $E(i)$, we can always
use a $K_4$ or $<i,j>$ or $E(i)$ to replace it \emph{validly}.
\end{lemma}

\textbf{Proof:} The proof is a direct derivation of Lemma
\ref{lemma2} because any block is a one-copy state. Using $K_4$ or
$<i,j>$ or $E(i)$ to replace $\ket{\psi}$ will guarantee the new set
$S$ contains $S_{\ket{\psi}}$. Therefore, we can replace
\emph{validly}.\hfill $\blacksquare$

Thus, directly by Lemma \ref{lemma2}, we will only consider $<i,j>$
or $K_4$ or $E(i)$ as blocks in following discussion. However, due
to the simplicity of analysis of the scheme when only $<i,j>$ type
states are in use, we seek to limit the number of other type states
in the scheme. Therefore, it is natural to add an additional
principle to the optimal scheme namely the more $<i,j>$ states in
use the better under the same number of copies. We can always find
such optimal scheme in all possible schemes with minimum number of
copies, which means adding the new principle will not change the
minimum number of copies for the problem. Finally, if we find in
certain scheme we can use $<i,j>$ states to replace other type
states \emph{validly}, such scheme must not be the optimal one in
our principle. We can treat the replacement in two different ways.
Firstly, a valid replacement is a indication that the current scheme
is not optimal and may be out of our concerns. Secondly, a valid
replacement can also be treated as a modification process to the
optimal scheme. Sometimes, when a replacement of more than one block
is necessary, the definition of validity is the natural extension of
the one block case. This concept is important to understand the
process of replacement in our following discussion.

Following the new principle, we want to obtain the property of the
optimal discrimination scheme. It is easy to see the number of
$E(i)$ in the optimal scheme is at most 1. Otherwise, say there are
two $E(i)$ and $E(j)$ in the optimal scheme, we can use $<i,j>$ and
$<i,k>$(any $k\notin\{i,j\}$) to replace them which contradicts its
optimality. We denote a discrimination scheme using only $<i,j>$ and
$K_4$ as limited scheme. Let the number of copies in the optimal
scheme be $t_{opt}$ and the number in the optimal limited scheme be
$t^{1}_{opt}$. If no $E(i)$ appears in the optimal scheme,
$t_{opt}=t^{l}_{opt}$. Otherwise, we can use $<i,j>$ and $<i,k>$(any
$k\notin\{i,j\}$) to replace $E(i)$ and obtain $t^{1}_{opt}\leq
t_{opt}+1$. Thus, $t_{opt}\geq t^{1}_{opt}-1$. The same idea will be
used again once we bound the number of $K_4$ in the optimal limited
scheme.

Next we shall seek for a lower bound of the limited scheme. For
description simplicity, we use a graph language to depict the
problem. We construct a graph $G$ for any limited scheme in the
following way. Each $<i,j>$ or $K_4$ used in the scheme is treated
as a type I or II vertex in the graph $G$ respectively. There are
only two types of edges in $G$. If any two type I vertexes have a
common element, there is an edge between them. For example, there is
an edge between $<i,j>$ and $<i,k>$.  If one type I vertex's
elements are included in a type II vertex, there is an edge between
them. For example, there is an edge between $<i,j>$ and
$K_4\{i,j,k,l\}$. It is easy and important to see for any $<i,j>$,
the pair $(i,j)$ can be and only can be discriminated by the vertex
adjacent to $<i,j>$.

It should be noticed that the graph here is a representation of the
discrimination scheme not the one we mention before to demonstrate
the power of discrimination of each type state. In $G$, we have $t$
vertexes, $l_1$ type I and $l_2$ type II as well as many connected
subgraphs $\{G_n\}$. There are $l_n$ vertexes in $G_n$ in which
$l_n^1$ type I and $l_n^2$ type II. Let $D(G')=\{d_i^j|
<d_i^1,d_i^2> \in G' \}\bigcup\{e_i^j|K_4\{e_i^1,e_i^2,e_i^3,e_i^4\}
\in G'\}$ for any graph $G'$. Then it is important to see the
following properties of the graph $G$.

\begin{lemma}\label{lemma3}\upshape
For any optimal limited discrimination scheme, the corresponding
graph $G$ has following properties. The degree\footnote{the degree
of a vertex is the number of edges whose one end is the vertex.} of
any type II vertex is 0. For any subgraph $G_n$ with at least one
type I vertex, namely only type I vertex due to the claim above,
$l_n\geq 2$ and $|D(G_n)|\leq l_n+1$.
\end{lemma}
\textbf{Proof.} First, we can easily obtain that the degree of any
type II vertex is at most $2$. Assume we have a type II vertex with
more than $2$ adjacent vertexes(type I). Because $K_4\{i,j,k,l\}$
has only $4$ distinct elements, at least 2 of its adjacent vertexes
have a common element, say $<i,j>$ and $<i,k>$. It is easy to see
that such a $K_4$ is unnecessary because it can not discriminate any
new pair. Thus, the only possible case where $K_4\{a,b,c,d\}$ has
degree 2 is its two adjacent vertexes must be $<a,b>$ and $<c,d>$.
However, we can replace $K_4\{a,b,c,d\}$ by $<a,c>$ in such case.
Therefore, degree $2$ is also impossible for any type II vertex.

The type II vertex with degree 1 shares an edge with a type I
vertex, say $<i,j>$ and $K_4\{i,j,k,l\}$, we can replace the $K_4$
by $<i,k>$. Therefore no type II vertex with degree $1$. Finally,
the degree of any type II vertex is $0$.

For any subgraph $G_n$ with at least one type I vertex, it can only
contain type I vertexes due to the result above. Any type I vertex
$<i,j>$ in $G_n$ implies a need of an adjacent vertex to
discriminate the pair $(i,j)$. Thus $l_n\geq 2$. Any edge implies
one repetition of appearance of the common element and there are at
least $l_n-1$ edges in $G_n$. Therefore, $|D(G_n)|\leq
2l_n-(l_n-1)=l_n+1$. \hfill $\blacksquare$

It is only the case $K_4$ with degree $0$  we have not discussed
yet. Denote the set of all such type II vertex as $G_K$. We have
following lemma to bound $|G_K|$. We define a pair is
\emph{uniquely} discriminated by a $K_4$ when no other vertex can
discriminate the pair.

\begin{lemma}\label{lemma4}\upshape
 $|G_K|\leq 9$.
\end{lemma}
\textbf{Proof.} Any $K_4$ in $G_K$ must discriminate at least one
pair \emph{uniquely}. Due to our principle of preferring $<i,j>$ to
$K_4$, it is easy to check any vertex in $G_K$ discriminates 1 or 2
pairs uniquely can always be replaced by $<i,j>$ state. For the
$K_4$ discriminating 3 pairs uniquely, only when it discriminates
the pair $(a,b)$,$(a,c)$ and $(b,c)$ (namely $K_3\{a,b,c\}$) it
cannot be replaced. However, we will prove such $K_4$ will appear at
most once. Once there are two such $K_4$, say $K_4\{a,b,c,d\}$ and
$K_4\{a',b',c',d'\}$, we can always replace them by two $<i,j>$ type
states. If $|\{a,b,c,d\}\cap\{a',b',c',d'\}|=3$, say $\{a,b,c\}$ are
common elements, we can replace them by $<a,b>$ and $<a,c>$.
Otherwise, $|\{a,b,c,d\}\cap\{a',b',c',d'\}|\leq2$, because two
$K_4\{a,b,c,d\}$ and $K_4\{a',b',c',d'\}$ discriminates
$K_3\{a,b,c\}$ and $K_3\{a',b',c'\}$ uniquely, thus
$|\{a,b,c\}\cap\{a',b',c'\}|\leq 1$ namely, at least two elements in
one $K_3$ are different from the corresponding ones in another
$K_3$, say $\{b, b',c,c'\}$ are distinct. It should be noticed that
to replace the two $K_4$, it is required that the new states can
discriminate the pairs which are discriminated \emph{uniquely} by
the set $\{K_4\{a,b,c,d\}, K_4\{a',b',c',d'\}\}$ in case there are
common pairs discriminated by both $K_4$. If
$|\{a,b,c,d\}\cap\{a',b',c',d'\}|\leq1$, there is no pair
discriminated by both $K_4$, then we can use $<b,b'>$ and $<c,c'>$
to replace the two $K_4$. If $|\{a,b,c,d\}\cap\{a',b',c',d'\}|=2$,
then $\{a,d\}$ are common elements, we use $<a,b>$ and $<b,b'>$
instead. Finally, we obtain that all $K_4$ but at most one in $G_K$
discriminate at least 4 pairs uniquely.

For any $K_4\{a,b,c,d\}$ in $G_K$, its 0 degree implies that once
any element say $a$ appears in any type I vertex, it must be $<a,e>$
where $e\notin \{a,b,c,d\}$. Namely, such $<a,e>$ discriminates 3
pairs in $\{a,b,c,d\}$ which makes $K_4$ fail to discriminate at
least 4 pairs uniquely because $K_4$ can discriminate 6 pairs in
total. Thus, all $K_4$ in $G_K$ do not share elements with other
type I vertex with at most one exception sharing 1 element.
Therefore at least $|D(G_K)|-1$ elements in $D(G_K)$ won't appear in
$D(G-G_K)$. Because any pair between these elements can only be
discriminated by the vertex in $G_K$. Then we have a necessary
condition
\begin{equation}
    \frac{1}{2}(|D(G_K)|-1)(|D(G_K)|-2)\leq 6|G_K|.
\end{equation}
All $K_4$ in $G_K$ can only discriminate pairs in $D(G_K)$, which
means all the pairs $G_K$ can discriminate won't exceed all the
pairs in $D(G_K)$. Thus we need $|G_K|\leq \frac{2}{3}|D(G_K)|+1$,
otherwise we can replace $G_K$ by the discrimination scheme proposed
at the beginning of the section using at most
$\frac{2}{3}|D(G_K)|+1$ copies. Combined with the inequality above,
we have $|D(G_K)|\leq 12$ and $|G_K|\leq \frac{2}{3}|D(G_K)|+1\leq
9$.\hfill $\blacksquare$

In order to obtain a lower bound for the scheme, we need to have a
necessary condition for a discrimination scheme as follows:

\begin{lemma}\label{lemma5}\upshape
For any limited discrimination scheme with corresponding graph $G$,
we have $|D(G)\cap\{1,...,N\}|\geq N-1$.
\end{lemma}
\textbf{Proof:} If there are at least two elements, say $a$ and $b$,
are not in any $<d_i^1,d_i^2>$ or $K_4\{e_i^1,e_i^2,e_i^3,e_i^4\}$,
then we can not discriminate the pair $(a,b)$ because this pair can
only be discriminated by $<a,c>$ or $<b,c> c\notin\{a,b\}$ state
where at least one of $\{a,b\}$ will appear or by $K_4\{a,b,c,d\}$
which both $\{a,b\}$ will appear. Therefore, there is at most one
element belonging to $\{1,...,N\}$ but not to $D(G)$. Finally, we
have $|D(G)\cap\{1,...,N\}|\geq N-1$.\hfill $\blacksquare$

\begin{theorem}\label{theorem1}\upshape
For any \emph{unentangled discrimination scheme} with $t$ copies, we
have $t\geq \frac{2}{3}N+o(N)$ asymptotically.
\end{theorem}
\textbf{Proof.} Because of Lemma \ref{lemma2}, we have the optimal
discrimination scheme can only use $<i,j>$ or $K_4$ or $E(i)$ state.
We further seek for a lower bound of limited discrimination scheme.
Due to Lemma \ref{lemma4}, $|G_K|\leq 9$, we can replace each
$K_4\{i,j,k,l\}$ in $G_K$ by $<i,j>$ and $<i,k>$ to make the scheme
use only type I vertexes , using extra $9$ copies. Using the same
idea when we obtain $t_{opt}\geq t^{1}_{opt}-1$, denote the number
of copies in the optimal scheme using only $<i,j>$ as $t^2_{opt}$,
we have $t^1_{opt}\geq t^2_{opt}-9$ and $t\geq t^2_{opt}-10$.

Then we want to obtain $t^2_{opt}$. The new corresponding $G$ has
only type I vertexes and the analysis in Lemma 3 is also valid, we
have $|D(G_n)|\leq l_n+1=\alpha_n l_n$ where $l_n\geq 2$. Thus
$\alpha_n\leq \frac{3}{2}$ and $|D(G_n)|\leq \frac{3}{2}l_n$.

\begin{equation*}
    |D(G)|=|\bigcup_{G_n}D(G_n)|\leq \sum_{G_n}|D(G_n)| \leq \sum_{G_n}
    \frac{3}{2}l_n=\frac{3}{2}t^2_{opt}
\end{equation*}

Because of Lemma 5, $|D(G)|\geq N-1$, we have $t^2_{opt}\geq
\frac{2}{3}(N-1)$ and finally $t\geq t_{opt} \geq
\frac{2}{3}N+o(N)$. \hfill $\blacksquare$

Finally, we prove $\frac{2}{3}N$ is the asymptotic lower bound of
the unentangled discrimination scheme. Combined with the
discrimination scheme proposed at the beginning of the section,
$\frac{2}{3}N$ is also a tight lower bound.

It is interesting to see that using such unentangled  discrimination
scheme we can solve the \emph{Grover's Oracle Identification
Problem} exactly using only $\frac{2}{3}N+o(N)$ queries, which is
less than $N-1$ queries by a classical algorithm. We do not use
entanglement which is considered to be the key role making quantum
computing superior to classical one in the scheme. On the other
side, only superposition and product state are used in the scheme.
In the next section we shall consider the discrimination scheme
where entanglement can be used.

\subsection{Lower bound for a general parallel discrimination scheme}
In order to achieve the general lower bound of such scheme, we first
deal with the general structure of the input state of $t$ copies
network. Recall the notation we use in ~\ref{eqn:expandoracle} and
~\ref{eqn:sgnoracle}, the general input state will be in following
form $ |\psi\rangle=\sum_{\vec{a}}p_{\vec{a}}|\vec{a}\rangle$. Let
$\sigma$ be a permutation on $\vec{a}$, it is easy to verify that
$\tau(\vec{a})=\tau(\sigma(\vec{a}))$. Therefore, it's reasonable to
consider $\vec{a}$ only by the number of each label.

For any $\vec{a}$, let $c_i$ be the number of $a_k$ such that
$a_k=i$, i.e., $c_i=|\{a_k|a_k=i,1\leq k\leq t\}|$. Obviously we
have $\sum_i c_i=t$. Assume there are $l_1$ odd elements and $l_2$
even elements in $c_i$. For odd $c_i$,  we have $c_i\geq 1$. Thus
$l_1\leq t$.

\begin{lemma}\label{lemma6}\upshape
Assume $t\leq \frac{N}{2}$. For any $\vec{a}$ with $\{c_i\}$
expression, the number of pairs $(i,j)$ which make $\tau(\vec{a})$
odd, denote as $n_{\vec{a}}$, is at most $t(N-t)$.
\end{lemma}
\textbf{Proof.} For any $\vec{a}$ with corresponding $\{c_i\}$ and
any pair $(i,j)$, it is obvious that $\tau(\vec{a})=c_i+c_j$. Thus,
the total number of pairs $(i,j)$ that satisfy $\tau(\vec{a})=1$ is
$l_1(N-l_1)$. Noting that $l_1\leq t$ and $t\leq \frac{N}{2}$, we
have $n_{\vec{a}}\leq t(N-t)$, the equality holds when $l_1=t$.
\hfill $\blacksquare$

The discriminating condition in Eq. (\ref{eqn:discondition}) here
becomes, for any pair $(i,j)$, $\sum_{\tau(\vec{a})=0}
|p_{\vec{a}}|^2-\sum_{\tau(\vec{a})=1} |p_{\vec{a}}|^2=0$. Because
$\sum_{\vec{a}}|p_{\vec{a}}|^2=1$, we have \\
\begin{equation}\label{eqn:multicondition}
    \sum_{\tau(\vec{a})=1} |p_{\vec{a}}|^2=\frac{1}{2}.
\end{equation}

\begin{theorem}\upshape
For any parallel discrimination scheme, the minimal number of copies
for perfectly identifying Grover Oracle is not less than
$\frac{1}{2}(N-\sqrt{N})$.
\end{theorem}

\textbf{Proof.} Without loss of generality, we may assume $t\leq
N/2$. Otherwise the result automatically holds. Summing up
Eq.~(\ref{eqn:multicondition}) for all pairs $(i,j)$, we have:
\begin{equation}\label{eqn:sumup}
\max_{\vec{a}}(n_{\vec{a}}) \sum_{\vec{a}}|p_{\vec{a}}|^2\geq
\sum_{\vec{a}} n_{\vec{a}} |p_{\vec{a}}|^2 =\frac{1}{4}N(N-1).
\end{equation}
If $t\leq \frac{N}{2}$, then it follows from Lemma \ref{lemma6} that
$\max_{\vec{a}}(n_{\vec{a}})\leq t(N-t)$. Thus we have
\begin{equation}
t(N-t)\geq \frac{1}{4}N(N-1),
\end{equation}
which together with the assumption $t\leq N/2$ implies that $t\geq
\frac{1}{2}(N-\sqrt{N})$. \hfill $\blacksquare$

Although we don't know whether the lower bound in the above theorem
is tight, we do have found some interesting examples (in the next
section) where the use of entanglement can dramatically reduce the
number of the queries and meet this lower bound.  So we believe that
$\frac{1}{6}N$ queries is likely be saved by employing entanglement.

It has been shown that computing the boolean function OR is closely
related to the unsorted database search problem \cite{WL07}.
Clearly, an algorithm for computing OR function can also be used to
do exact quantum search. The converse part, however, is not
necessarily true. The complexity of computing OR function is in
general higher than that of the exact quantum search. Our results
are helpful in understanding such a difference. More precisely,
in~Ref. \cite{BBCM+98} it has been shown that $N$ quires is a tight
lower bound for computing OR function, but here we have shown that
$\frac{2}{3}N+1$ is an upper bound for exact quantum search even in
the absence of entanglement, and it is likely that such an upper
bound can be reduced with the assistance of entanglement.
\subsection{Some examples}
We shall present several example to demonstrate our results. We have
analyzed the cases of $N=2,3,4$ when we discuss the power of
one-copy state. In this section we present two additional examples
to demonstrate our results.

\begin{example}\label{example1}\upshape
It is easy to verify that when $N=5$, it is impossible to identify
unknown orale with certainty by just one single use. Interestingly,
there does exist a product discrimination scheme using just two
queries. The input state is given as follows:
\begin{equation}\label{eqn:entangleexample1}
\ket{\Phi_5} =E(1)\otimes K_4\{2,3,4,5\},
\end{equation}
where
$E(1)=\frac{1}{\sqrt{3}}\ket{1}+\frac{1}{\sqrt{6}}(\ket{2}+\ket{3}+\ket{4}+\ket{5})$
and $K_4\{2,3,4,5\}=\frac{1}{2}(\ket{2}+\ket{3}+\ket{4}+\ket{5})$.

The validity of the above scheme can be verified directly. Another
way to see this is that $K_5$ is just covered by
$S_{K_4\{1,2,3,4\}}$ and $S_{E(5)}$. We should point out that $2$
queries is less then half of $5$. That means its power beyond
$\frac{1}{2}N$ lower bound.\hfill $\blacksquare$
\end{example}

However, we cannot solve the case of $N=6$ using just two queries
without the assistance of entanglement. Remarkably,  by using an
entangled state as input, we can achieve a perfect identification
for the case $N=6$.
\begin{example}\label{example2}\upshape
Take
\begin{equation}\label{entangled-6}
\ket{\Phi_6}=\frac{1}{4}({\sum_{1\leq i<j\leq 6}\ket{ij}+\ket{33}}),
\end{equation}
where $\ket{33}$ can be replaced by any $\ket{kk}$ such that $1\leq
k\leq 6$. We shall show that $f_k^{\otimes 2}\ket{\Phi_6}$ should be
mutually orthogonal. A simple argument is as follows. It is clear
that each $f_k^{\otimes 2}\ket{\Phi_6}$ contains exactly five terms
with $``-"$ sign. Taking  inner product between $f_k^{\otimes
2}\ket{\Phi_6}$ and $f_l^{\otimes 2}\ket{\Phi_6}$, the sign before
$\ket{kl}$ is changed into $``+"$, which results in a summation with
$8$ $``-"$ signs. It follows from the lower bound
$\frac{1}{2}(N-\sqrt{N})$ that two queries are optimal.\hfill
$\blacksquare$
\end{example}

\subsection{Solution to the general case}
In the above, we completely analyzed the \emph{Grover's Oracle
Identification Problem} which is a particular case of the unitary
operation discrimination problem. Although the solution to the most
general problem remains open, we can still make use of our
techniques in the unentangled scheme to analyze some more general
problems.

In our technique above, the \emph{distinction graph} plays a very
important role to understand the power of a state to discriminate
unitary operations pair in the unentangled scheme. Also it is
important for the possibility to employ graph and combinatorial
skills in our analysis. For more general problem, such a
\emph{distinction graph} will also be helpful for our analysis.

One generalization of the current problem is to generalize the form
of the oracle function $f$. Namely, $f_i$ can be any function
mapping $\{1..N\}$ onto $\{0,1\}$ and the quantum oracle is still
the same form as the one in Eq.~\ref{oracle-simplified}. In such
case, the \emph{distinction graph} and the latter combinatorial
skills can still provide much help because of the fine discrete
structure of such problems. However, when general unitary operations
are considered, the \emph{distinction graph} method might not be
very efficient and powerful as before.

\section{Summary} \label{sec:summary}
In conclusion, we generalize the unsorted database search problem to
the \emph{Grover's Oracle Identification Problem} which reveals both
algorithmic and unitary operations distinguishing aspect of the
problem and show the connection between them.  In analysis, we
obtain a tight lower bound $\frac{2}{3}N$ for the product
discrimination scheme and a general lower bound
$\frac{1}{2}(N-\sqrt{N})$ for the parallel discrimination scheme.
Finally, we also show that the complexity of exact quantum search
with one unique solution can be strictly less than that of the
calculation of OR function. Some interesting examples are also
presented. Further more, we provide a brief idea to the solution
with more general case.

There are still many interesting unsolved problems. For instance, we
would like to know whether $\frac{1}{2}(N-\sqrt{N})$ is a tight
lower bound for such scheme. Second, the lower bound for general
unitary operation identification problem seems to be a great
challenge.

We are indebted to the colleagues in the Quantum Computation and
Quantum Information Research Group for many enjoyable conversations.
In particular, we sincerely thank Prof. Mingsheng Ying for his
numerous encouragement and constant support on this research. This
work was partly supported by the National Natural Science Foundation
of China (Grant Nos. 60702080, 60736011, 60503001, and 60621062),
the FANEDD under Grant No. 200755, and the Hi-Tech Research and
Development Program of China (863 project) (Grant No. 2006AA01Z102).

\end{document}